\documentclass[
    twocolumn,
    superscriptaddress,
    amsmath,amssymb,
    pra,
    aps,
    floatfix,
    longbibliography,
]{revtex4-2}

\usepackage{dcolumn}
\usepackage{physics}
\usepackage[caption=false, position=top]{subfig}
\usepackage{tikz}
\usepackage{xparse}
\usepackage{graphicx}
\usetikzlibrary{quantikz}
\usepackage{xurl}
\usepackage[shortlabels]{enumitem}
\usepackage{bm}
\usepackage[inkscapearea=page]{svg}

\NewDocumentCommand\DownArrow{O{2.0ex} O{black}}{%
    \mathrel{\tikz[baseline] \draw [<-, line width=0.5pt, #2] (0,0) -- ++(0,#1);}
}

\begin{document}

\preprint{APS/123-QED}

\title{Hybrid quantum optimization in the context of minimizing traffic congestion}

\author{Jedwin Villanueva}
\email{jedwinv@unimelb.edu.au}
\affiliation{School of Physics, University of Melbourne, Parkville, VIC, 3010, Australia.}
\author{Gary J Mooney}
\email{mooneyg@unimelb.edu.au}
\affiliation{School of Physics, University of Melbourne, Parkville, VIC, 3010, Australia.}

\author{Bhaskar Roy Bardhan}
\affiliation{Research and Advanced Engineering, Ford Motor Company, Dearborn, MI 48124, USA.}
\author{Joydip Ghosh}
\affiliation{Research and Advanced Engineering, Ford Motor Company, Dearborn, MI 48124, USA.}
\author{Charles D Hill}
\affiliation{School of Physics, University of Melbourne, Parkville, VIC, 3010, Australia.}
\affiliation{School of Mathematics and Statistics, University of Melbourne, VIC, Parkville, 3010, Australia.}
\author{Lloyd C L Hollenberg}
\email{lloydch@unimelb.edu.au}
\affiliation{School of Physics, University of Melbourne, Parkville, VIC, 3010, Australia.}


\date{\today}

\begin{abstract}
Traffic optimization on roads is a highly complex problem, with one important aspect being minimization of traffic congestion. By mapping to an Ising formulation of the traffic congestion problem, we benchmark solutions obtained from the Quantum Approximate optimization Algorithm (QAOA), a hybrid quantum-classical algorithm. In principle, as the number of QAOA layers approaches infinity, the solutions should reach optimality. On the other hand, short-depth QAOA circuits are known to have limited performance. We show that using tailored initialization techniques encourages the convergence to the desired solution state at lower circuit depths with two and three QAOA layers, thus highlighting the importance of adapting quantum algorithms in the noisy intermediate scale (NISQ) quantum computing era. Moreover, for NISQ devices with limited qubit connectivity and circuit depth, we introduce a heuristic noise-resilient variant of QAOA predicated on the elimination of long-range 2-qubit interactions in the QAOA layers whilst the cost function is unaltered. Our results show that this QAOA variant is surprisingly effective, outperforming QAOA on a physical IBM Quantum computer device.
\end{abstract}

\maketitle


\section{\label{sec:intro}Introduction}

As quantum computer technology continues to traverse the noisy intermediate scale quantum (NISQ) regime towards full-scale error-corrected systems~\cite{Preskill_2018, Xu2023, Kim2023}, the implementation of quantum algorithms for real-world applications such as optimization has received much attention~\cite{9504923, math10030416, Salehi2022}. The benchmark approach is the quantum approximate optimization algorithm (QAOA)~\cite{QAOA} typically applied to problems encoded in a quadratic unconstrained binary optimization (QUBO) formulation or equivalently an Ising formulation~\cite{ising-formulation}. QAOA has also shown promise in addressing binary optimization problems with higher order terms~\cite{sachdeva2024quantum}.
%
%
%
It remains to be seen whether such generic quantum optimisation approaches in this nascent NISQ era can compete with highly developed classical heuristics that have enjoyed many decades of evolution, e.g. for the travelling salesman problem (TSP)~\cite{johnson_mcgeoch_1995, Johnson2007, 10.1007/3-540-52846-6_97, HELSGAUN2000106, 10.1145/290179.290180, Applegate2003ChainedLF} as well as more general QUBO problems~\cite{Dunning}. Nevertheless, the implementation, benchmarking and improvement of quantum optimization approaches for real-world problems on NISQ devices is a topic of great interest~\cite{Brandhofer2022, Barkoutsos_2020, PhysRevA.104.012403, Khumalo2022}. Within these confines, a key question focuses on improving implementation of quantum optimization algorithms on realistic hardware in the NISQ era, potentially informing how such calculations might be better performed on full-scale error corrected systems. In this paper we investigate implementations of the QAOA approach in the context of an important real-world problem – minimization of traffic congestion. Through numerical simulation in the first instance, we study the effect of initialization and the use of precomputed parameters to avoid barren plateau issues in the classical minimization steps. The effect of noise on the solutions was studied via implementation on IBM Quantum devices, and we introduce a hardware-informed QAOA ansatz compression strategy to also avoid noise. Under different metrics that quantify the quality of solutions and runtime, we examine how QAOA solutions scale over a range of problems of up to 23 cars. 

Since the proposal of QAOA, the performance and limitations of the algorithm have been intensively studied in general classes of problems which are typically in the form of MaxCut~\cite{PhysRevA.103.042612, Guerreschi2019, 8957201, PhysRevLett.125.260505, PhysRevLett.124.090504, Alam2019AnalysisOQ}. Generally, the quality of the QAOA solution scales with the circuit depth, along with the number of variational parameters. One of the main challenges come from the noise-free barren plateau (BP) problem~\cite{McClean_2018} which ultimately makes the classical optimization of the QAOA circuits NP-Hard~\cite{Bittel_2021}. The noise-induced barren plateau (NIBP) is a similar BP phenomenon causing vanishing gradients in the presence of noise~\cite{Wang}, further obscuring the global optimum during classical optimization. To alleviate the difficulty of parameter optimization, various initialization techniques have been proposed and investigated~\cite{QAOA-Zhou_2020,TQA-Sack_2021, multistartQAOA}, including the use of nearly-optimal precomputed parameters~\cite{Brando2018ForFC, fixed-angle-QAOA, transfer_median, Galda2021TransferabilityOO, QAOA-ML}. To further reduce the classical burden, we investigate the algorithm's performance using precomputed parameters to omit the optimization of the QAOA algorithm. This approach largely avoids the BP and the NIBP problems. Other methods of omitting the optimization are possible for short-depth QAOA circuits via analytical means~\cite{Ozaeta_2022, Hadfield_2023}, as well as for QAOA circuits with low connectivity via reverse causal cones which provide ideal measurements via classical simulation of reduced quantum circuits~\cite{Streif_2020}.

Despite rapidly improving quantum hardware~\cite{PhysRevLett.127.180501, Arute2019, Mooney2019, Mooney_2021, Mooney2021_2}, another key factor to the performance of QAOA on a quantum processor is the noise induced by large-depth circuits, restricting the size of reliable quantum computation. Even a single layer of QAOA on a general QUBO instance can result in densely connected QAOA circuits. Running such circuits requires the use of SWAP gates in order to realize the interactions of non-connected qubits on the hardware. This impacts heavily on the performance~\cite{Lotshaw2022}, prompting a surge of research aiming to improve or compress the QAOA ansatz~\cite{fromqaoatoqaoa_Hadfield, Egger_2021, PhysRevResearch.4.033029, 10.1145/3549554, Cook_2020, Bartschi_2020, QAMPA, MQAOA} and implement advanced error mitigation techniques~\cite{Kim2023, sachdeva2024quantum}. Here, we introduce and investigate a heuristic for reducing noise on physical hardware by removing 2-qubit interaction gates that require SWAP gates. As this fundamentally changes the underlying variational ansatz, we re-optimized the compressed circuits with respect to the original problem Hamiltonian while introducing extra variational parameters.

This paper is organized as follows: in Section II, we introduce the traffic problem and its mapping to QUBO form. In Section III, we benchmark baseline QAOA, approaches to initialization, and the use of precomputed QAOA parameters. Results from running on hardware are given in Section IV. In Section V we implement connectivity-forced QAOA ansatz compression tailored to the hardware as well as various measures of run time for comparison analyses. Conclusions are given in Section VI. 


\section{\label{sec:background} Background }
\subsection{\label{sec:problem_statement} Traffic Problem Statement}

\begin{figure*}[!htbp]
    \centering
    \includegraphics[width=0.9\textwidth]{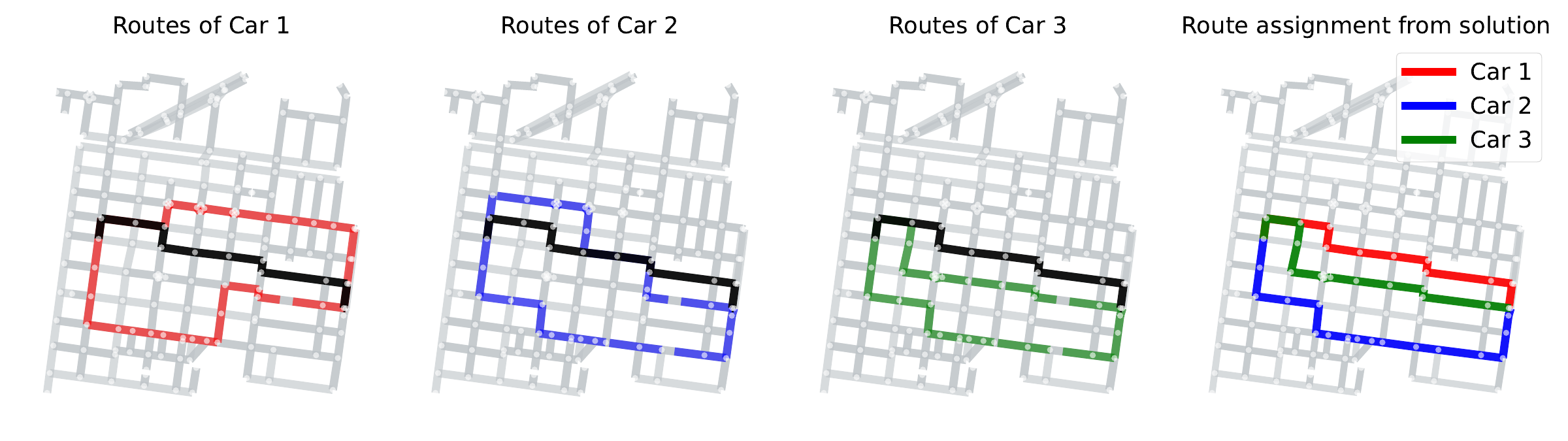}
    \caption{A problem of 3 cars, 3 routes each (9 variables) showing the designated routes. The black route is a common route between all cars chosen to be the shortest route. The dashed lines show two dissimilar, alternative routes (detours). A solution to the QUBO is plotted in the rightmost panel, showing minimal traffic congestion. Provided the cars are traveling in the same direction, the problem can be interpreted as depicting either direction of travel.}
    \label{fig:trafficproblem}
\end{figure*}

The problem of reducing congestion on roads can be modeled as a binary combinatorial optimization problem by defining decision variables corresponding to the route taken by each car, as previously done in~\cite{Neukart}. Given a set of cars, we assigned each car a set of possible routes to travel from their origin to destination. A route is a path consisting of edges corresponding to road segments. Each route begins and ends respectively at origin and destination nodes, which were always chosen to be located at intersections for simplicity. An example problem is shown in Fig \ref{fig:trafficproblem}. The decision variables $q_{ij}$ for car $i$ taking route $j$ can be assigned as,

\begin{equation}
\label{eqn:binary_vars}
    q_{ij} = \left\{
        \begin{array}{ll}
        1, & \text{if car $i$ taking route $j$ is True}\\
        0, & \text{otherwise.}
        \end{array}
        \right.
\end{equation}
 
A cost is assigned to each segment, depending on which cars have routes that travel on that segment. Namely, for the $k$-th road segment in a road network, labelled $s_k$, the cost is the sum
\begin{equation}
\label{eqn:segment_cost}
    c(s_k) = d_k\cdot{\left(\sum_i\sum_{j\in R_i}{q_{ij}}\right)}^2.
\end{equation}
where $R_i = \{j |$ car $i$ route $j$ contains $ s_k \}$ is the set of routes for car $i$ that contain segment $s_k$, with $d_k>0$ a positive weight for the segment, such as the segment's physical length. The prefactor \(d_k\) can be set to 1 if all segments are equally weighted. If multiple cars travel on a segment $s_k$, then $c(s_k)$ will increase quadratically with the number of non-zero binary variables $q_{ij}$. This disincentivises multiple cars traveling on the segment. The total traffic congestion contribution to the cost over the network can be taken as, 
\begin{equation}
    \label{eqn:total_segment_cost}
    A = \sum_{k}c(s_k).
\end{equation}
minimizing this term corresponds to minimizing the congestion over the whole road network. However, this cost function is trivially minimized by setting every variable to zero. We thus enforce appropriate constraints stating that each car must take exactly one route. The constraint for each car takes the form
\begin{equation}
    \label{eqn:single_car_constraint}
    B_i = 1-\sum_{j}{q_{ij}} = 0,
\end{equation}
noting that the sum is over all routes $j$ of car $i$, so that only one $q_{ij}=1$ within the set of routes of a car. The constraints can be incorporated into the the cost function by minimizing $(B_i)^2 > 0$. The total constraint contribution to the cost function can be obtained by summing over all cars,
\begin{equation}
    \label{eqn:car_constraint}
    B = \sum_{i}{(B_{i})^2} > 0.
\end{equation}
The traffic problem can now be formulated as a quadratic unconstrained binary optimization (QUBO) problem using the cost function, 
\begin{equation}
    \label{eqn:objective}
    C = A + \lambda B,
\end{equation}
where the factor $\lambda > 0$ needs to be sufficiently large to penalize violation of constraints. As a heuristic choice, we set $\lambda$ so that it scales the range of possible cost values of $\lambda B$ to be equal to the range of $A$. That is, $\lambda$ is chosen such that 
\[ {\max}_{q_{ij}}(A) - {\min}_{q_{ij}}(A) = {\max}_{q_{ij}}(\lambda B) - {\min}_{q_{ij}}(\lambda B),\] using the observation
\begin{align*}
    \label{eqn:minmaxcostvalues}
    \mathrm{argmin}_{q_{ij}}(\lambda B) = \mathrm{argmin}_{q_{ij}}(A) &= \{0, 0, ... 0\} \text{ and, } \\
    \mathrm{argmax}_{q_{ij}}(\lambda B) = \mathrm{argmax}_{q_{ij}}(A) &= \{1, 1, ... 1\}.
\end{align*}
We found that this method of choosing $\lambda$ resulted in robust solutions that successfully satisfied the constraints in all of our experiments.


\subsection{\label{sec:mapping_to_qaoa}Mapping to QAOA}

The first step in mapping the QUBO problem to the QAOA is to transform the cost function of binary variables to a quantum Ising Hamiltonian consisting of Pauli-Z operators $Z_{u}$ with eigenvalues $z_{u} \in \{-1, 1\}$. The mapping is given by, 
\begin{equation} 
    \label{eqn:variable_mapping}
    q_{ij} \mapsto \frac{1}{2}(1-Z_{u}),
\end{equation}
where $u$ represents a re-indexing of the pairs of indices $(i, j)$, ordered in ascending order with $j$ increasing first and then $i$. Under this mapping, the QUBO cost function is transformed to the Ising Hamiltonian
\begin{equation}
    \label{eqn:Ising_Hamiltonian}
    H_C = \sum_{uv}{J_{uv}}Z_u Z_v + \sum_{u} h_u Z_u + c,
\end{equation}
up to a scalar constant $c$. The Ising Hamiltonian can be visually represented as a graph, where edges are weighted according to $J_{uv}$ and nodes are weighted according to $h_u$.
The expectation values of the Ising Hamiltonian with respect to a superposition of quantum states give the average cost of those states weighted by the state probabilities. By minimizing the expectation value with respect to a variational ansatz, it is possible approximate the ground state of $H_C$, resulting in a quantum state that maximises the probability of the ground state. The QAOA ansatz is defined as,
\begin{equation}
    \label{eqn:trial_state}
    \ket*{\Psi(\vec{\gamma}, \vec{\beta})} = \displaystyle \prod_{l=1}^{p}{\displaystyle e^{\displaystyle-i\beta_l H_B} \displaystyle e^{\displaystyle-i\gamma_l H_C}} \ket{+}^{\otimes N},
\end{equation}
where $\vec{\gamma} = (\gamma_1, \gamma_2, ...)$ and $\vec{\beta} = (\beta_1, \beta_2, ...)$ are the variational parameters, and $N$ is the total number of qubits. $H_B = -\sum_u X_u$ is a mixer Hamiltonian consisting of Pauli-X operators summed for each qubit $u$ for which the initial state \(\ket{+}^{\otimes N}\) is the ground state of $H_B$. The two unitary operators in each of the $p$ factors in the product are time-evolution operators of $H_C$ and $H_B$. The initial state \(\ket{+}^{\otimes N}\) can be prepared on a quantum computer as
\begin{equation}
    \label{eqn:initial_state}
    \ket{+}^{\otimes N} = H^{\otimes N} \ket{0}^{\otimes N}.
\end{equation}
The variational method is applied by minimizing the expectation value of the Ising Hamiltonian  
\begin{equation}
    \label{eqn:matrix_element}
    \expval{H_c}=\mel*{\Psi(\vec{\gamma}, \vec{\beta})}{H_c}{\Psi(\vec{\gamma}, \vec{\beta})}
\end{equation}
with respect to the variational parameters $\vec{\gamma}$ and $\vec{\beta}$. A representative quantum circuit of the QAOA ansatz is shown in Fig \ref{fig:QAOASchematic}. 
The Ising Hamiltonian can be normalised by some non-zero positive constant which is equivalent to a re-scaling of the parameter vector \(\vec{\gamma}\). We perform normalization such that the average absolute values of coefficients, \(J_{uv}\) and \(h_u\) is equal to 1. This was done so that \mbox{\(-\pi \le \gamma_i \le \pi\)} is a suitable search space, since values outside this range are likely to be too large and structure from the problem Hamiltonian encoded in the circuit is dominated. Without rescaling, the parameter landscape exhibits the narrow basin of attraction in~\cite{transfer_median}, where optimization features are narrow and is thus difficult for traditional optimizers. We also use the translational symmetry \(\beta_i \mapsto \beta_i \pm \pi\) which leaves the ansatz unchanged in order to restrict \(-\pi/2 \le \beta_i < \pi/2\).

\section{\label{sec:lit_review_qaoa}QAOA initialisation}
The performance of QAOA is dependent on the variational parameters that define the final QAOA circuit. However, finding the optimal QAOA parameters can be computationally difficult, with studies showing that the classical search for optimal parameters is in fact NP-Hard~\cite{Bittel_2021} via gradient descent methods with randomly initialized QAOA parameters. We implemented various initialization techniques from literature, and benchmarked their performance on the traffic problems against random initialization.

Some of the first initialization techniques that were proposed involve incrementally adding a QAOA layer, using the optimal parameters found from the $p$-layer QAOA to give an initial guess to the $(p+1)$-layer QAOA. This means that at each step, the number of variational parameters increases by 2. We considered two methods based on this approach called INTERP and FOURIER~\cite{QAOA-Zhou_2020}. INTERP uses the observation that the optimal $p$-layer QAOA parameters tend to follow the same scheduling as the optimal $(p+1)$-layer QAOA, thus an initial guess to the $(p+1)$-layer QAOA is a linearly interpolation from the optimal $p$-layer QAOA parameters. FOURIER is a re-parametrisation of the variational parameters via a discrete sine/cosine transformation,
\begin{align}
    \label{eqn:FOURIER}
    \gamma_i &= \sum_{k=1}^{q}{u_k\sin{\left( \left[k-\frac{1}{2}\right]\left[i-\frac{1}{2}\right]\frac{\pi}{p} \right)}},\\
    \beta_i &= \sum_{k=1}^{q}{v_k\sin{\left( \left[k-\frac{1}{2}\right]\left[i-\frac{1}{2}\right]\frac{\pi}{p} \right)}},
\end{align}
optimizing over the FOURIER parameters $\vec{u}$ and $\vec{v}$ instead. The $(p+1)$-layer QAOA parameters is initialized by using the same FOURIER parameters of the optimal $p$-layer QAOA. optimization is also performed with respect to the FOURIER parameters. Another promising initialization technique we considered is Trotterised Quantum Annealing (TQA)~\cite{TQA-Sack_2021}, utilising the connection between QAOA and quantum annealing. In particular, the continuous time evolution of a linear-ramp quantum annealing schedule is discretised via Trotterisation, resulting in the alternating unitary operators of QAOA and a formula for the QAOA parameters given by,
\begin{equation}
    \label{eqn:TQA_params}
    \gamma_j = \frac{j}{p}\Delta t,\ \beta_j = (1-\frac{j}{p})\Delta t,
\end{equation}
which involves the Trotter time step $\Delta t$ as a parameter. The Trotter time step is related to the total quantum annealing time $T$ and the number of QAOA layers $p$, as $T=p\Delta t$. The optimal value for $\Delta t$ is not known and varies between different QAOA instances. Similar to the presence of QAOA parameter concentration - which has been observed on certain problem classes as instance sizes increase~\cite{brandao2018fixed, Akshay_2021} - it was observed that the best values for $\Delta t$ also concentrate~\cite{TQA-Sack_2021}. In particular, an optimal concentration value of $\Delta t \sim 0.75$ was found for typical weighted and unweighted MaxCut QAOA instances.

\subsection{\label{sec:performance_metric}Performance Metric: Approximation Measure}
We want a way of measuring the quality of the QAOA quantum states \( \ket{\psi}=\ket*{\Psi(\vec{\gamma},\vec{\beta})} \), as well as a single computational basis state \(\ket{\psi}=\ket{z}\). We define the instance specific approximation measure
\begin{equation}
    \label{eqn:approx_quality}
    R_{\rm true} := \frac{ E_{\max} - \expval{H_C} }{ E_{\max} - E_{\min} }, 
\end{equation}
where \(E_{\max}\) is the maximum energy eigenvalue of \(H_C\), \(E_{\min}\) is the minimum energy eigenvalue, and \(\expval{H_C} = \mel{\psi}{H_C}{\psi}\) is the expected energy with respect to the target quantum state. The lowest energy state corresponds to \(R_{\rm true}=1\), and the highest energy state corresponds to \(R_{\rm true}=0\). For larger problems, \(E_\mathrm{max}\) and \(E_\mathrm{min}\) is classically difficult to calculate. To reduce computational demands, we use a metric that substitutes \(E_{\min}\) with a close approximation of the ground state energy and \(E_{\max}\) with the mean energy derived from a uniform distribution of states. We define this as the approximation measure \emph{relative to random},

\begin{equation}
    \label{eqn:rand_approx_quality}
    R_{\rm random} := \frac{ E_{\rm random} - \expval{H_C} }{ E_{\rm random} - \overline{E}_{\min} } 
                    = \frac{ c - \expval{H_C} }{ c - \overline{E}_{\min} }.
\end{equation}

where \(E_\mathrm{random}\) is the average energy of the \(2^N\) possible states. To obtain the simplified equation on the right of Eq \eqref{eqn:rand_approx_quality}, we note that \(E_\mathrm{random}\) can be calculated as,
\begin{align}
    E_\mathrm{random} &= \mel{+^N}{H_C}{+^N}, \text{ where }
    \ket{+^N} := \ket{+}^{\otimes N},
\end{align}
and since \(\mel{+^N}{Z_i}{+^N}=0\) and \(\mel{+^N}{Z_i Z_j}{+^N}=0\), the contributions of the Pauli terms of \(H_C\) vanish, leaving only the scalar constant \(c\) in Eq \eqref{eqn:Ising_Hamiltonian}. To approximate \(\overline{E}_{\min}\), we use a state-of-the-art commercial solver, Gurobi ~\cite{gurobi}, which is highly likely to be optimal for the small problem instances in this work. In summary, a positive value \(R_{\rm random} > 0\) represents an expected energy better than randomly guessing, and a value close to \(1\) indicates near-optimality. While the initial measure \( R_{\rm true} > 0 \) for any possible quantum state, the adjustment to \( R_{\rm true} \) means that this new measure can be negative for some states for which its average performance is worse than randomly guessing.

\subsection{\label{sec:experimental_setup}Numerical Experiment Setup}
\begin{figure*}[!htbp]
    \centering
    \includegraphics[scale=0.75]{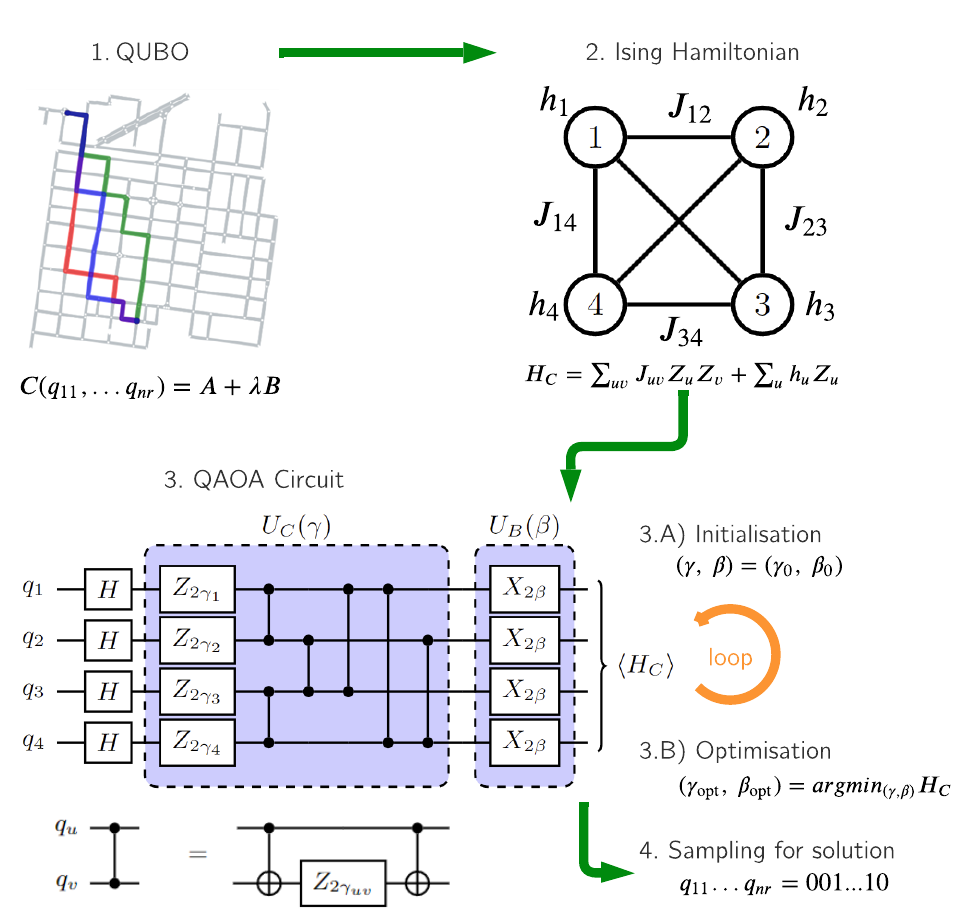}
    \caption{Schematic of the QAOA algorithm for the traffic congestion problem. 1) An instance of the traffic problem is formulated as a QUBO problem. 2) The QUBO is mapped to a quantum Ising Hamiltonian by mapping binary variables to Pauli-Z operators, such that the problem solution is encoded as the ground state. 3) A parameterized QAOA circuit is constructed using the Ising Hamiltonian, consisting of layers of alternating unitaries \(U_C(\gamma)\) and \(U_B(\beta)\). The decomposition of the ZZ-rotation gates is shown below. The angles for the single-qubit gates are shown as subscripts where \(\gamma_u := \gamma h_u\) and \(\gamma_{uv} = \gamma J_{uv}\). 3) The QAOA classical optimization loop occurs by repeatedly measuring the energy and classically updating the parameters, represented by the circular arrow. 3A) The parameters are initialized. 3B) The classical optimization loop updates the parameters \( (\gamma, \beta) \) until they converge \( (\gamma_\mathrm{opt}, \beta_\mathrm{opt}) \), minimizing the expectation of the Hamiltonian. 4) The optimized QAOA state is sampled from the quantum device to find an approximate solution to the original problem.}
    \label{fig:QAOASchematic}
\end{figure*}

In the following, traffic problems of various numbers of cars and routes were randomly generated using the Open Street Map Python package called OSMnx~\cite{osmnx_Boeing_2017} which was used to obtain road distances of a region within the city of Melbourne, Australia. For a given number of cars and number of routes, we fixed the starting and destination nodes which were the road intersections and used a k-shortest paths algorithm to generate 1000 routes for each instance. For each car we then randomly selected routes according to a uniform distribution, in addition to the shortest possible path. Once the required number of routes were selected for each car, the traffic instance was formulated as a QUBO problem and mapped to QAOA. The QAOA circuits were then simulated without noise using Qiskit~\cite{Qiskit}, an open-source quantum framework. The following shows an outline of the preparation of the traffic instances and QAOA solutions, as schematically shown in Fig \ref{fig:QAOASchematic}.

\begin{enumerate}
    \item \emph{QUBO Formulation}. Prepare multiple routes from an origin node to a destination node. Randomly assign to each car a number of routes (see Fig 1 for an example). Define the QUBO objective function to minimize.
    \item \emph{Convert to Ising Formulation}. Use the mapping described in Eq\eqref{eqn:variable_mapping} to obtain the Ising Hamiltonian in Eq\eqref{eqn:Ising_Hamiltonian}. 
    \item \emph{Construct $p$-layer QAOA Ansatz}. Construct the variational state in Eq\eqref{eqn:trial_state} on a quantum computer. The final QAOA circuit is then obtained by the following two steps.
    \begin{enumerate}[A)]
        \item \emph{Initialize $2p$ QAOA parameters}. initialize QAOA parameters for the QAOA circuit optimization. We used random initialization and three existing techniques found in the literature on variational quantum algorithms.
        \item \emph{Classical optimization}. Use a classical optimizer to find the optimal QAOA parameters such that the ansatz minimizes the expected energy of the Ising Hamiltonian. We used an off-the-shelf gradient descent optimizer, SciPy's L-BFGS-B, however most optimizers should be sufficient.
    \end{enumerate}
    \item \emph{Measure the QAOA State}. Repeatedly run and measure the optimized QAOA circuit on a quantum computer, obtaining a distribution of binary strings. There are two choices when taking a single solution state from the distribution. Under the assumption that the QAOA state prepares the ground state with high probability, the first option is to take the most occuring string in the distribution. The second option is to evaluate the cost of each binary string and choose the lowest energy string at the cost of extra classical processing. We consider both solutions in Section \ref{sec:hfqaoa}.
\end{enumerate}

We also investigated the performance of using precomputed initial parameters without the classical optimization in Step 3B. As a practical consideration, this avoids the BP and NIBP problems, and reduces the number of quantum circuit executions required. Methods used to calculate optimal precomputed parameters are generally computationally intensive, however this overhead can be considered worthwhile. Once the parameters are optimized, they can frequently be applied to multiple graphs with similar properties~\cite{fixed-angle-QAOA}. Additionally, methods exist to adapt these parameters to achieve near-optimal results for graphs with varied structure~\cite{transfer_median, Galda2021TransferabilityOO}.

In order to generate precomputed parameters for our traffic problems, first we generated traffic problems of varying sizes. We then measure the mean and standard deviations of the Ising coefficients resulting from those traffic problems. We then randomly generated 100 Ising Hamiltonians of 9-qubits with coefficients $J_{uv}$ and $h_u$ chosen from a normal distribution with the same mean and standard deviation calculated from the generated traffic problems. Since the traffic problems typically result in fully connected Ising Hamiltonian graphs attributing to the constraint terms, we also used completely connected Ising Hamiltonians for this step. We then perform and optimize the QAOA circuits, and used the median of those 100 optimized parameters as the precomputed parameters of all the traffic problems in our numerical experiments.

\subsection{\label{sec:numerical_results}Numerical Results: QAOA Benchmark}

\subsubsection{\label{sec:random_vs_TQA}RANDOM and TQA Initialization}

The importance of initialization techniques is investigated by benchmarking TQA initialization against a naive technique of randomized (RANDOM) initialization, where QAOA parameters were drawn at random from a uniform distribution. The performance of each initialization technique was benchmarked for 9-qubit instances of the traffic congestion problem with 3 cars and 3 routes, with \(p=2\) and \(p=3\) layers of QAOA. We initialized 50 QAOA parameters for each approach, totaling 100 initializations. For RANDOM, the QAOA parameters were drawn from the ranges \(\gamma_i \in [-\pi, \pi)\) and \(\beta_i \in [-\pi/2, \pi/2)\). For TQA, the Trotter time steps $\Delta t$ were 50 equally spaced values in the range \(\Delta t \in [0.1, 1)\). The minimum value of \(\Delta t = 0.1\) avoids initializing values which are too small which is equivalent to setting \( \gamma_i = 0, \beta_i = 0\) in the optimization. Each RANDOM and TQA initialized QAOA parameters are then optimized via Scipy's L-BFGS-B optimizer with a fixed maximum number of 150 iterations, which was sufficient for convergence for both initialization approaches. 

\begin{figure*}[!p]
    \fontsize{8}{10}
    \centering
    \includegraphics[width=0.95\textwidth]{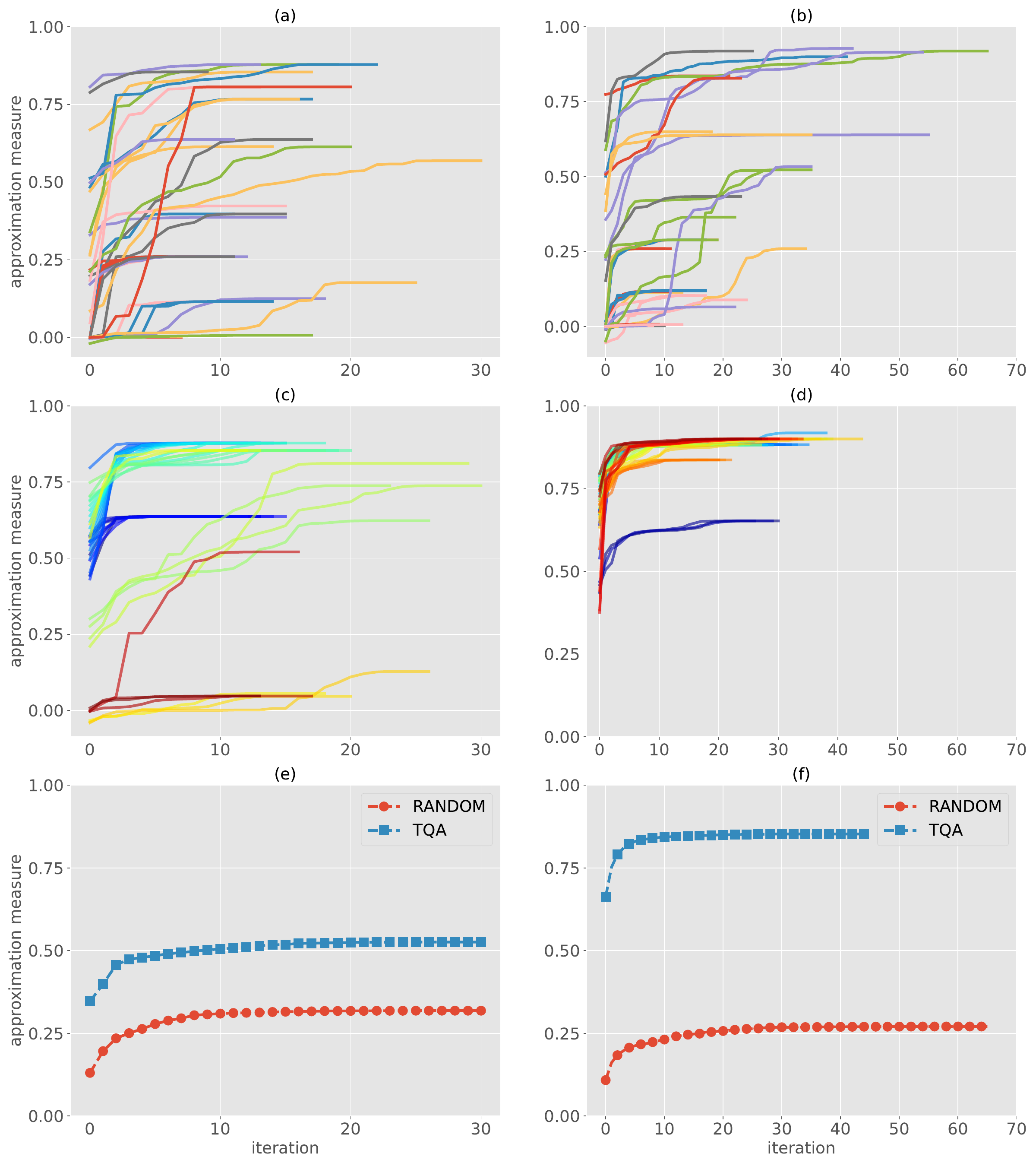}
    \caption{ Plots for a single traffic problem instance that show the convergence of the approximation measure for the evolving QAOA states during the optimization process using 50 QAOA parameters from (a)(b) RANDOM and (c)(d) TQA initialization techniques, and (e)(f) comparisons in performance between the two initialization techniques calculated as the average approximation measure over the corresponding 50 initializations. The line colors distinguish between different initializations. The left and right columns indicate values for $p=2$ and $p=3$ QAOA layers respectively. }
    \vspace{2cm}
    \label{fig:convergence_approx_RIvsTQA}
\end{figure*}

\begin{figure*}[!htbp]
    \fontsize{8}{10}
    \centering
    \includegraphics[width=0.95\textwidth]{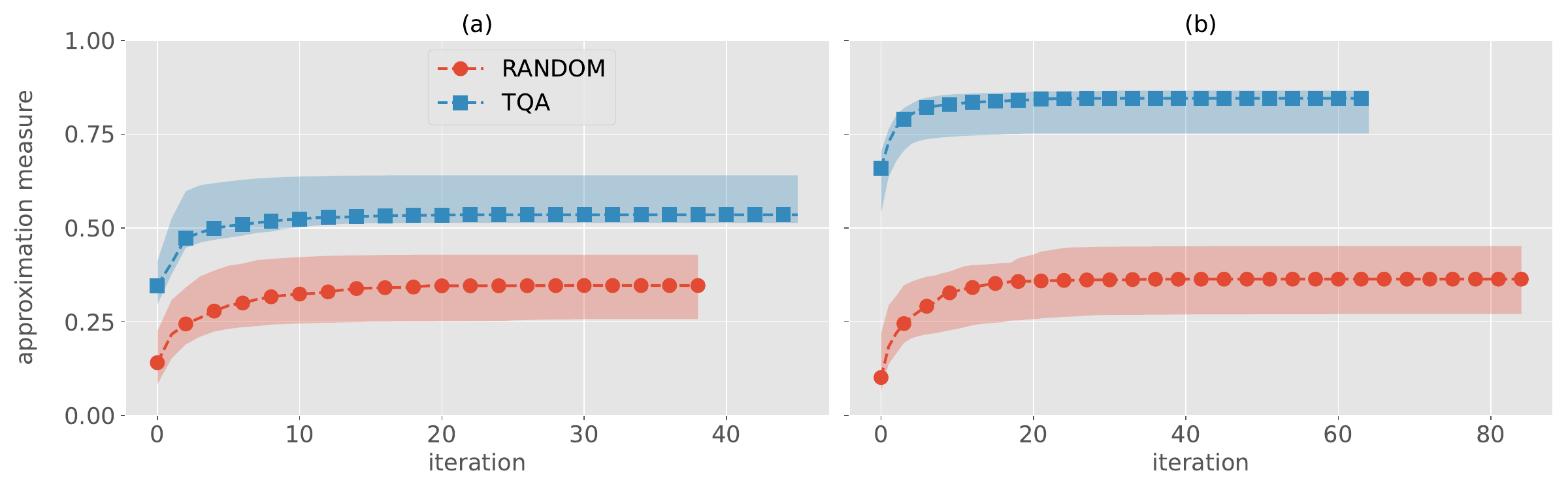}
    \caption{ Plots of comparisons of the RANDOM and TQA initialization techniques for (a) $p=2$ and (b) $p=3$ QAOA layers, taking the average approximation measure over the corresponding 50 initializations for each problem instance. The shaded region shows the range of recorded values for each technique.}
    \label{fig:convergence_approx_RIvsTQA_avg_2}
\end{figure*}

We plot the averages over 50 initialized parameters over all 20 traffic instances in Fig \ref{fig:convergence_approx_RIvsTQA_avg_2}. The shaded regions show the range between minimum and maximum average convergence, and the solid line corresponds to the median traffic instance. Across the 20 traffic instances, it is clear that TQA initialization outperforms random initialization on average. In fact, this suggests that randomly guessing \(\Delta t\) is better than randomly guessing the parameters. We also plot the convergence of each parameter for a single traffic instance in Fig \ref{fig:convergence_approx_RIvsTQA}, where each run is colored to distinguish between the 50 different initialized parameters. For this instance, TQA had more parameters which converged to a higher approximation measure. By observing the averaged approximation measures over the 50 initialized parameters, the TQA initialization method shows a distinct advantage in solution quality over RANDOM. The difference is more significant with \(p=3\) layers of QAOA.

\subsubsection{\label{sec:performance_initialized}Performance of Initialization Strategies}
We observed that initializing parameters with tailored techniques performs well even without optimization. This is particularly useful for precomputed parameters, which may be obtainable through noiseless simulations by looking at smaller instances or sub-graphs of ensembles of traffic instances, justified from parameter concentration in MaxCut instances independent of problem size~\cite{brandao2018fixed, Akshay_2021}.

We benchmarked the performance of initial parameters against the optimal parameters obtained after optimization using \(p=2\) QAOA. For INTERP and FOURIER, the optimal parameters for \(p=1\) were initially calculated via a global grid search, followed by gradient descent to further refine the optimal parameters. Then, the $p=2$ parameters for INTERP and FOURIER were calculated from the optimal $p=1$ parameters. Additionally, we performed gradient descent from the precomputed parameters to obtain the optimal parameters for $p=2$ QAOA. For each of the parameters, we took the probability distributions of the QAOA states to construct a density plot showing the approximation measure vs probability of a single state. In an ideal scenario where \(r_q = 1\) for the QAOA state, this would correspond to a 100\% probability of measuring the ground state, and hence a singular point at \( (Prob = 1, r_q = 1) \). However, this is generally not the case, especially for low-depth QAOA circ\textbf{}uits. Instead better performance is indicated by a greater probability of measuring states with higher approximation measures, which would correspond to higher concentrations of data-points on the top-right of the plots.

\begin{figure*}[!htbp]
    \fontsize{8}{10}
    \centering
    \includegraphics[width=0.9\textwidth]{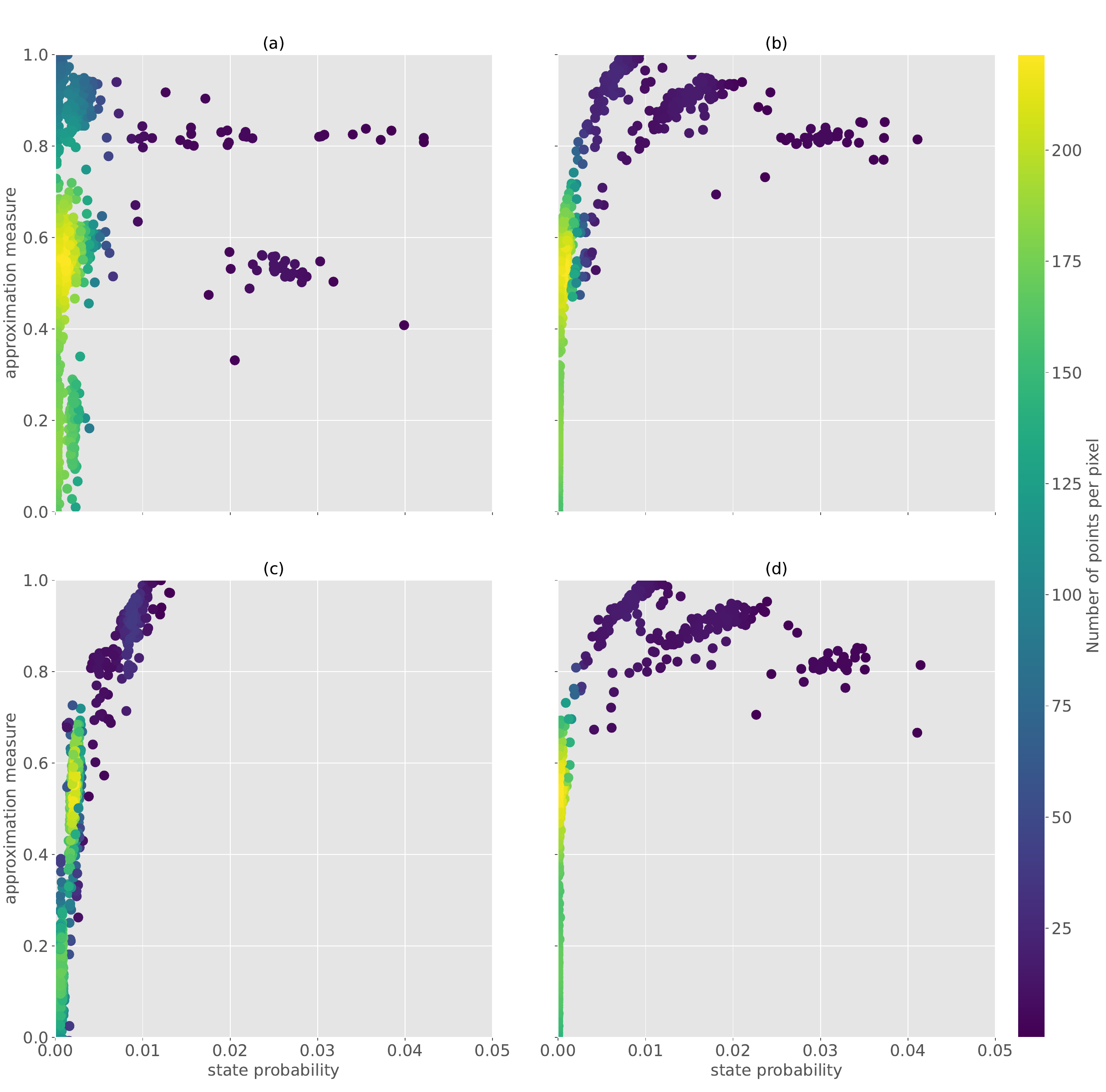}
    \caption{Density plots of binary states from the QAOA($p=2$) state probability distribution parameterised with (a) INTERP, (b) FOURIER, (c) precomputed \emph{initial} parameters, and (d) \emph{optimal} parameters, plotting the probability and approximation measures of each binary state. The optimal parameters are optimized with Scipy L-BFGS-B, initialized with the precomputed parameters. By superimposing 20 9-qubit traffic instances, there are $20\times2^9=10240$ data points for the $2^9=512$ possible binary states of each instance, with brighter regions showing higher concentration of points. The red dotted vertical line shows the background probability of $1/2^9\approx0.002$,  equivalent to sampling the QAOA initial state $\ket{+}^{\otimes N}$.} 
    \label{fig:FOURIER_INTERP}
\end{figure*}

The density plots of the 20 9-qubit traffic instances are plotted in Fig \ref{fig:FOURIER_INTERP}. For each instance, the optimal QAOA state always had the lowest average energy even though the highest probability state is generally not the ground state. In the plots, this corresponds to the states with \(y = 0.8\sim 0.9 \) approximation measures having the highest probabilities \(x \ge 0.03\). For some instances, there are states with approximation measures as low as 0.6 with high probability. Interestingly, FOURIER is most similar to the optimized state, but the reason for this is not explored due to the practical consideration that FOURIER still requires classical optimization of the $p-1$-layer QAOA while precomputed parameter initialization does not. In contrast, precomputed parameters seem to have the most reliable performance between all instances, as it tends to have the ground state with highest probability than other states. INTERP is the least performing method, suggesting that it is likely not useful for directly sampling the QAOA state without optimization.

\section{\label{sec:real_results}NISQ Hardware results}

\subsection{\label{sec:runtime}Performance of QAOA on NISQ hardware}
Here we investigated the expected QAOA runtime scaling and naively compare with runtime scaling of the Gurobi classical solver, which use highly complex mathematical models to solve such integer problems~\cite{gurobi}. Knowing the challenges of noisy optimisation due to NIBPs, we performed noiseless optimization of the QAOA circuits via simulations. The pre-optimised QAOA circuits are then executed on real quantum processors from IBMQ. This limits the destructive impact of noise observed in our numerical experiments to the final QAOA circuits only. We also observed that in most cases, precomputed parameters perform almost as well as optimized parameters on noisy quantum computers, since the prepared state is likely close to a mixed state. Thus, using pre-optimised parameters from noiseless simulations is valid in this noisy regime under the assumption of the availability of precomputed parameters, which is heavily discussed in previous sections. Additionally, to minimize circuit depths, we only perform QAOA with $p=1$ layers. In order to see how performance scales with increasing problem size, we incremented the size of the traffic problems by adding a car with two possible routes each, beginning from a trivial case of 1 car with 2 routes. Each increment increases the qubit requirement by 2.

\begin{figure}
    \centering
    \fontsize{8}{10}
    \includegraphics[width=0.45\textwidth]{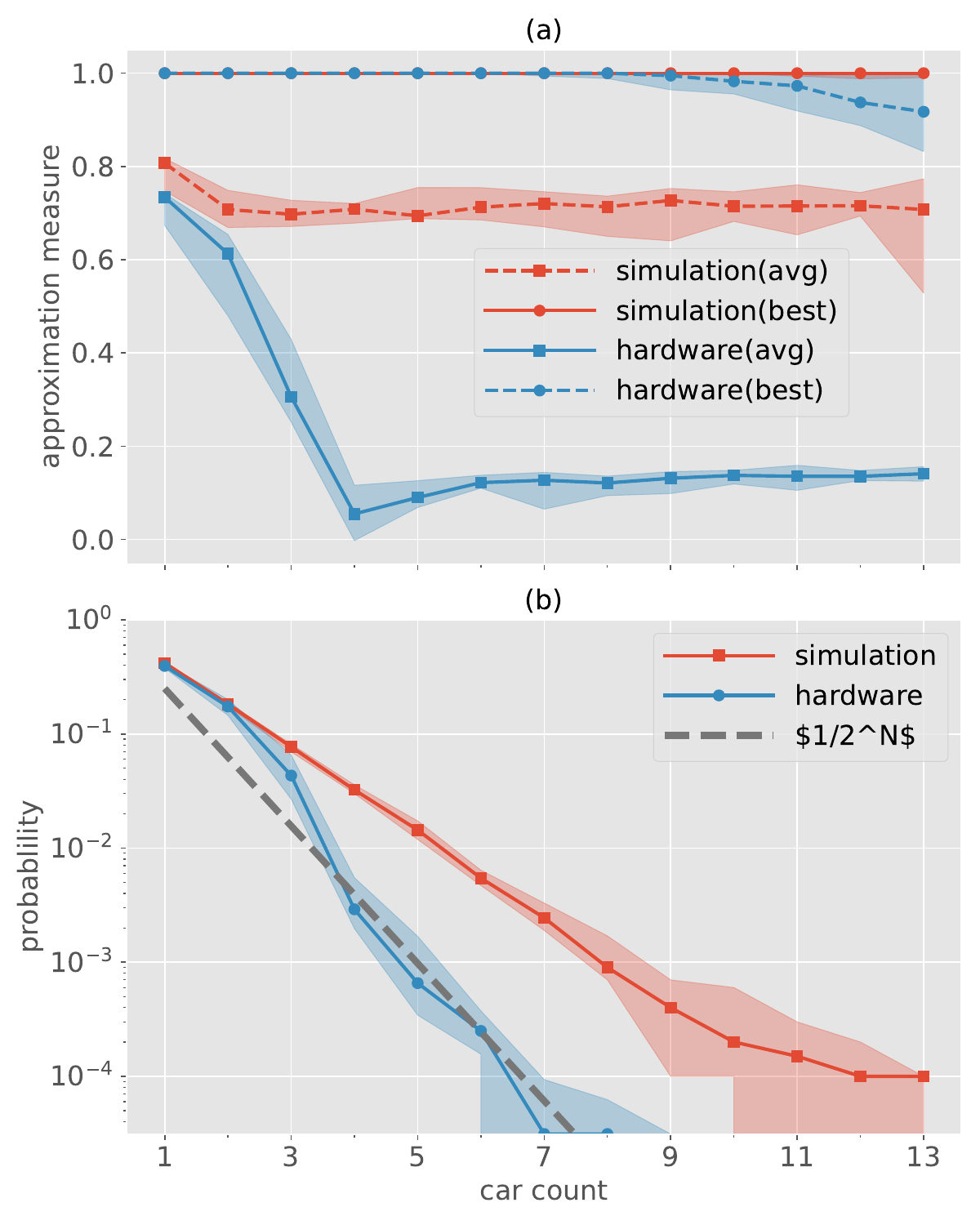}
    \caption{Results with $10000$ shots using the noiseless Qiskit Aer simulator (matrix product state) labeled \emph{simulation}, and $32000$ shots on IBM Quantum ibmq\_montreal labeled \emph{hardware}. The plots show the (a) approximation measure and (b) probability of the ground state against increasing problem size. We use two values for the approximation measure. The first is the expected value over QAOA distribution labeled \emph{avg}(blue), and the second is the approximation measure of the \emph{best} solution state(orange) with the lowest energy in the distribution. In all cases, there were 2 routes assigned to each car, thus the number of variables/qubits is double the number of cars. Each point shows the median of 10 problem instances, and the shaded region is the range of all recorded values over the 10 problem instances.}
    \label{fig:QAOAsimulationdata}
\end{figure}

We first present the results of QAOA on simulators vs real devices to show how impact of noise on performance. In Fig \ref{fig:QAOAsimulationdata}, we plot the approximation measure on the left, as well as the probability of the ground state on the right, against the number of cars. For the approximation measure, we use two values, the average approximation measure over the whole QAOA distribution, labeled \emph{avg}, and the maximal approximation measure of any one state in the distribution, labeled \emph{best}, corresponding to the single best performing state observed in the QAOA distribution. The numerical experiment was performed on both ibmq\_qasm\_simulator (simulation) and ibmq\_kolkata (hardware).

In the noiseless case, we see that the approximation measure of QAOA is \(\sim\)1 for the best solution state, and \(\sim\)0.7 for the average over the distribution. Unsurprisingly, the performance of the QAOA state significantly deteriorated on the noisy hardware. This is likely attributed to the increased circuit depths and CNOT counts of the densely connected QAOA circuits when transpiled to the physical IBM Quantum's ibmq\_kolkata with a heavy hex architecture~\cite{Qiskit}(Fig. 4). The average approximation measure plateaus at some positive value, suggesting that even as the problem size increases, the QAOA ditribution will output a solution better than random guessing on average. Furthermore, the best solution state sampled from the noisy QAOA circuit reports a value no smaller than 0.8. However, when looking at the observed probability of the ground state, we see an exponential decrease in the noiseless case, but it is not as fast as the exponential increase of the search space dimension \(2^N\). Note that for problems with more than 11 cars (22 qubits), there were already instances where the ground state is not measured from the 10000 shots, suggesting more shots are required to measure the ground state. In the presence of noise, the rate of decrease in probability is much worse, first approaching the exponential rate of the search space dimension before dropping to a zero probability for instances with more than 10 qubits (5 cars). This means that the output solution of such an algorithm will almost always never be the exact solution for some large enough problem size with only a fixed number of shots. This is in contrast to classical algorithms, namely Goemans-Williamson with polynomial time scaling and a guarantee of \emph{minimum} performance. Instead, the noisy QAOA(p=1) state has a limit on the \emph{maximum} performance of the output state.

\section{\label{sec:hfqaoa}Connectivity Forced QAOA Ansatz}

\begin{figure}[!htbp]
    \centering
    \includegraphics[width=0.45\textwidth]{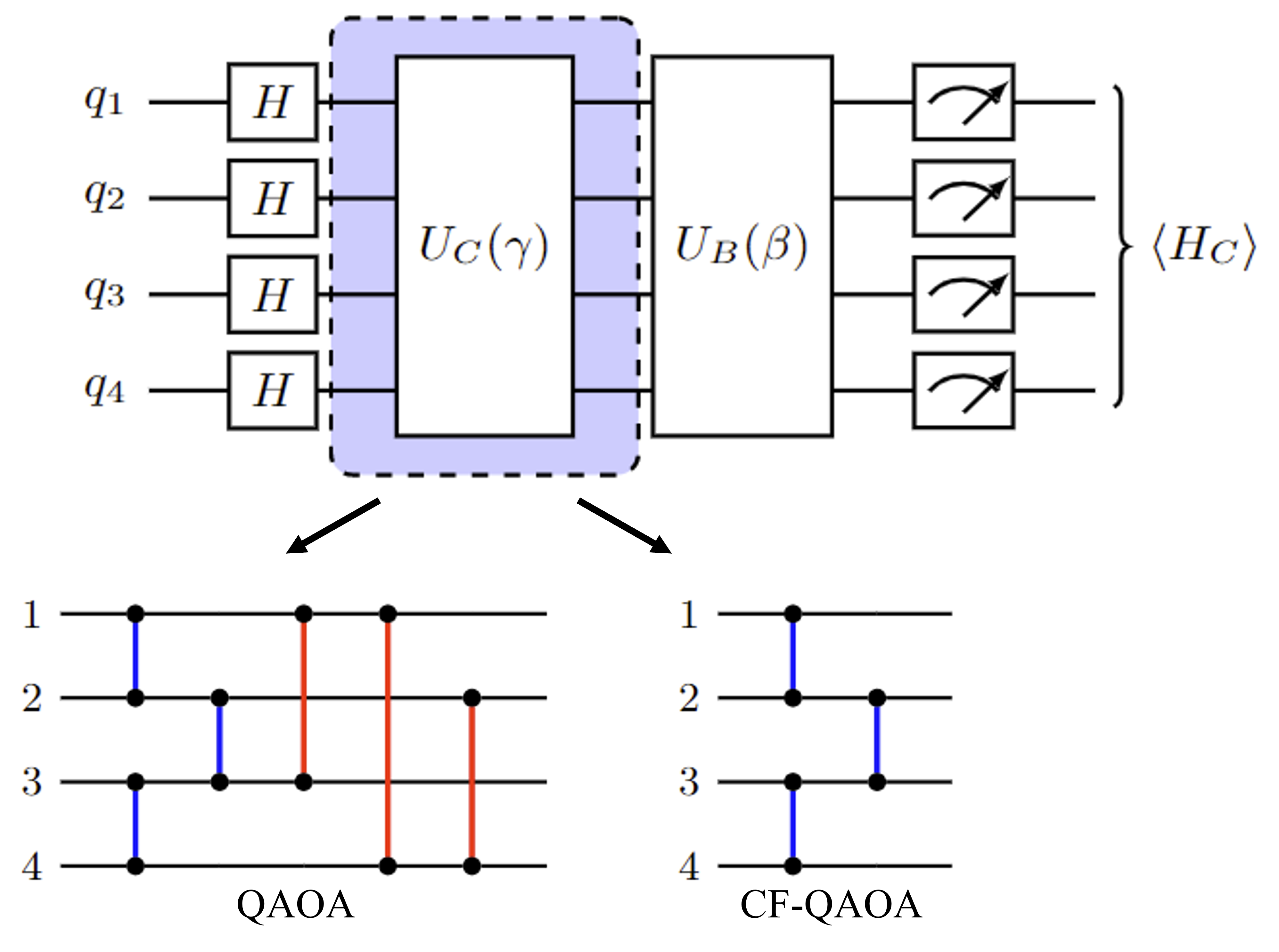}
    \caption{The modified unitary of the connectivity forced QAOA circuits (CF-QAOA), showing the removal of the red RZZ gates from the original QAOA circuits to implement on a 1d linear array of physical qubits.} 
    \label{fig:NewQAOASchematic}
\end{figure}

Since the previous section shows the destructive impact of noise, we heuristically counteracted noise in the simplest way possible by removing all RZZ gates that will induce SWAP gates according to the connectivity of the hardware, as in Fig \ref{fig:NewQAOASchematic}. On its own, such a method is ill-defined since there are many ways to map the problem variables to the physical qubits on the device, with each different mapping resulting in different RZZ gates on the hardware. Nevertheless, we will show that such a modification to circuits will result in better performing solutions under the restrictions of noise and limited connectivity on the hardware. The circuits with excised gates are referred to as connectivity-forced QAOA (CF-QAOA) circuits. The removal of the RZZ gates are done to entirely remove the necessity of SWAP gates, which is expected to destroy most quantum properties of the QAOA circuit. The CF-QAOA circuit itself is a QAOA circuit containing layers of RZ, RZZ, and RX gates in alternating fashion however it will correspond to a different underlying Ising Hamiltonian with less ZZ terms. Nevertheless, we treat CF-QAOA as a variational ansatz with respect to the original problem Hamiltonian and as such, optimisation is performed by minimising the expectation including all the ZZ terms from the original problem.

From the variational principle, removing any gate from an optimal QAOA circuit will result in a higher energy state, since this will change the prepared state. To counteract this increase in energy, we introduce extra variational freedom in the circuit without introducing additional gates by assigning the remaining rotation gates their own variational parameters in the CF-QAOA circuit in the same way that the multi-angle QAOA (ma-QAOA)~\cite{MQAOA} does for QAOA. We refer to these circuits as \emph{CF-maQAOA}. Additional variational parameters could mean that the optimization landscape is susceptible to more local minima. More sophisticated analysis would be required to precisely determine the classical burden added by the additional variational parameters. For QAOA and CF-QAOA, the number of parameters is always \( 2 p \), independent of the problem size. However, in ma-QAOA, the number of parameters also depends on the problem size per layer, increasing up to \(O(p (N^2+N))\) corresponding to the $O(N^2)$ maximal scaling of the number of terms in an Ising Hamiltonian. However, for CF-maQAOA, the number of parameters in each layer is bounded by the qubit connectivity. In the case of the heavy hex architecture, CF-maQAOA scales only linearly in the problem size, scaling as \(O(p (2.5N+N)) \) assuming an average connectivity of \(d=2.5\). We used the optimized QAOA parameters to initialize both CF-QAOA and CF-maQAOA circuits.

\subsection{Circuit Depths}
\begin{figure*}[!htbp]
    \centering
    \fontsize{8}{10}
    \includegraphics[width=0.95\textwidth]{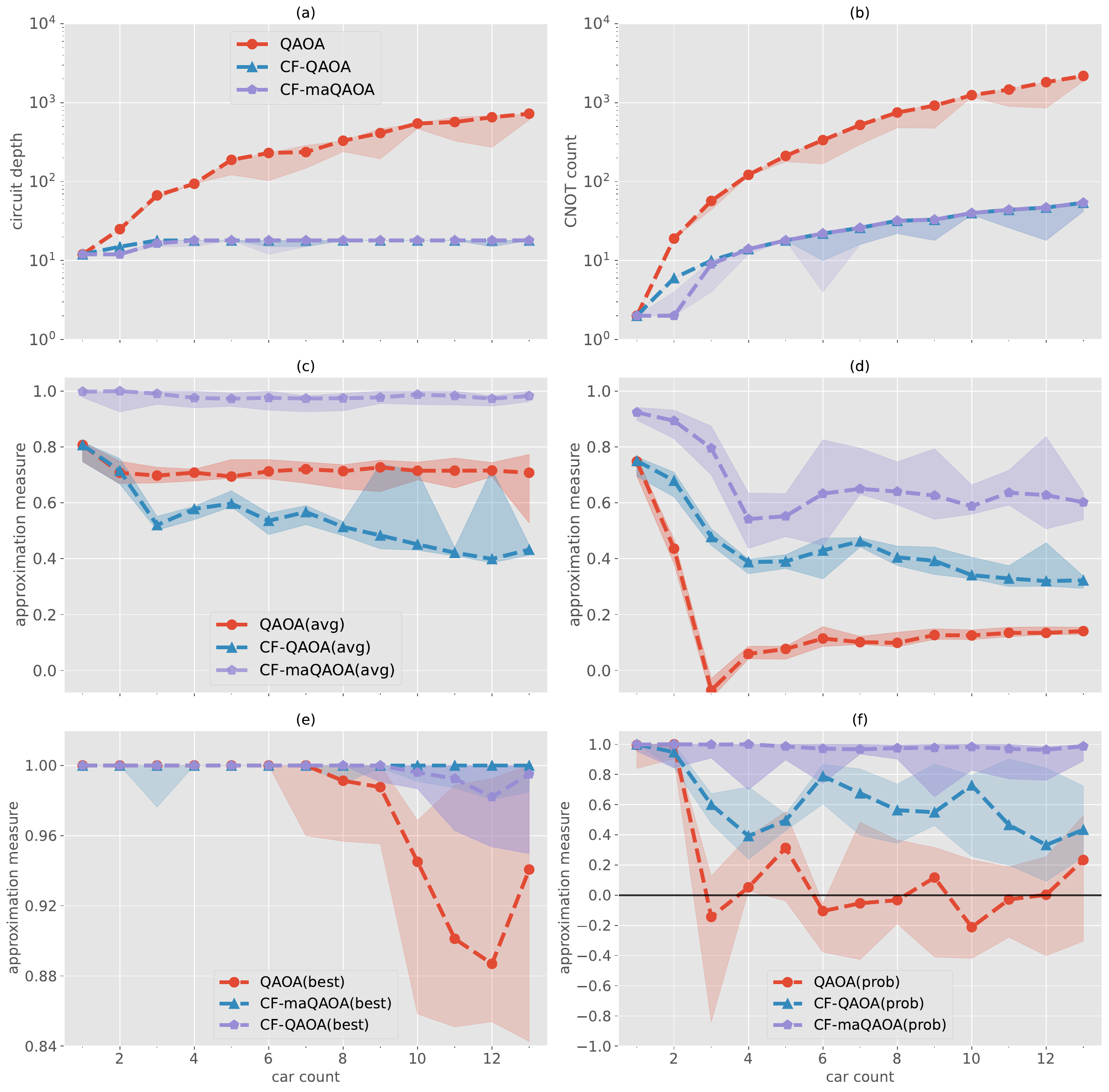}
    \caption{The transpiled (a) circuit depth and (b) CNOT counts for QAOA, CF-QAOA and CF-maQAOA implementations. Average approximation measure of three QAOA implementations from (c) noiseless simulations, and (d) physical hardware on \emph{ibmq\_kolkata}. The approximation measure of (e) the single best solution state and (f) the most probable solution state obtained from the QAOA distributions with 32000 circuit executions. For approximation measure, we plot horizontal lines for maximal approximation measure \(y=1\) and random guessing approximation measure \(y=0\) as benchmarks. We remark that dotted lines indicate a \emph{noiseless} simulation whereas solid lines indicate results from a physical device.  The shaded region indicates the whole range of values over 10 traffic instances and the marked point indicate the median value.}
    \label{fig:combined}
\end{figure*}

As in Section \ref{sec:real_results}, we performed optimization without noise and compared the performance of QAOA, CF-QAOA and CF-maQAOA running both on noiseless simulators and real hardware. We also performed readout error mitigation on the noisy circuits using Mthree~\cite{Mthree}. We first report the circuit depths and CNOT counts in the QAOA circuits after transpiling to the heavy hex architecture, shown in Fig \ref{fig:combined}(a)(b). Both CF-QAOA and CF-maQAOA reach a constant depth in contrast to a continually increasing circuit depth for QAOA. The rate of increase for the number of CNOTs is also much slower for both the CF-QAOA and CF-maQAOA.


\subsection{Performance of Modified Circuits}
Next, we report our numerical and empirical results for the average approximation measures of QAOA, CF-QAOA and CF-maQAOA in Fig \ref{fig:combined}(c)(d). Without noise in Fig \ref{fig:combined}(c), the average performance of CF-QAOA (orange) is worse than QAOA (blue). This is expected due to the damaging removal of the RZZ gates. However, on the noisy hardware, CF-QAOA outperforms QAOA. CF-maQAOA (green) increases the performance in the noiseless case to almost ideal in the noiseless case, likely due to the additional variational freedom. In the noisy case of Fig \ref{fig:combined}(d), there is still improvement from the CF-QAOA performance but it is no longer close to ideal. CF-maQAOA is the best performing in both scenarios.

We examined the approximation measures of potential solution states that can be obtained from the distribution upon repeated executions of the circuit. We report two candidate solutions, the \emph{best performing state} in Fig \ref{fig:combined}(e) with the highest approximation measure of the realized distribution, and the \emph{most probable state} in Fig \ref{fig:combined}(f) with the most occurrences. The QAOA implementation resulted in a maximal performance of no less than 0.84, which agrees with previous results in Fig \ref{fig:QAOAsimulationdata}(a). For CF-QAOA, the maximal performance is no less than 0.96, and for CF-maQAOA it is 0.98. Using the most probable state, the results are again consistent from simulations in \ref{fig:QAOAsimulationdata}(b) where performance approaches that of random guessing. However, for CF-QAOA and CF-maQAOA, this is not the case, with better than random performance for the larger instances.

\subsection{Runtime}
\begin{figure*}[!htbp]
    \centering
    \fontsize{8}{10}
    \includegraphics[width=0.95\textwidth]{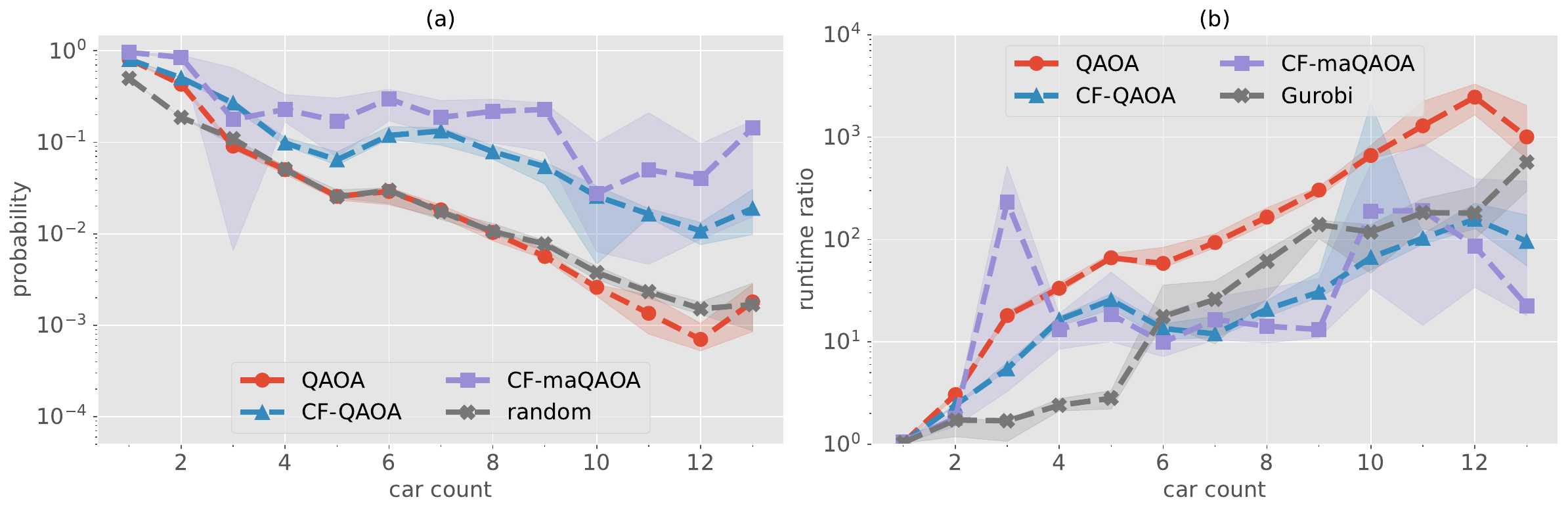}
    \caption{ In (a), the datapoints show the observed total probability of states with more than 0.8 approximation measure from the realized QAOA distributions executed on \emph{ibmq\_kolkata}. As a benchmark, we compare against random sampling of the equal superposition state. In (b) the ratios of the estimated runtime using the probabilities in (a) and, as a benchmark, the ratio of the exact recorded runtime of the Gurobi solver running on a single CPU of 3.10GHz clock speed. All ratios are normalized against the smallest recorded value within sample, thus each curve starts at $\sim 1$. Due to large variance between instances, we restrict the shaded region to the inter-quartile range (IQR) of instances for each car count for visual clarity. }
    \label{fig:combined_2}
\end{figure*}

Finally, we use the observed probabilities of the noisy QAOA distributions to estimate the expected runtime scaling. From previous considerations, we recorded vanishing ground state probability with a fixed number of shots as problem size increased. Thus, it is not possible to use these recorded probabilities of the ground state to estimate a runtime for general instances, as a zero probability corresponds to an infinite runtime. To circumvent this issue, we relax the constraint of requiring the exact ground state, and instead accept an approximate solution state with more than 0.8 approximation measure. Solution states with this lower bound are always obtainable in our experiments, as shown in Fig \ref{fig:combined}(e). Thus, instead of the ground state probability, the runtime measure is calculated based on the \emph{total probability} of states with an approximation measure greater than 0.8, plotted in Fig \ref{fig:combined_2}(a). Under this relaxation, we no longer use the dimension scaling of \(2^N\) to approximate the probability of randomly guessing the ground state as \(1/2^N\) but instead take the total number of states in the solution space with approximation measure greater than 0.8 divided by \(2^N\), labeled as \emph{random} in the plots. We see again that the total probability for QAOA on a physical device is scaling similarly to random, even with this relaxation of approximate solution states. CF-QAOA improves on this, and CF-maQAOA generally provides the most improvement but with more variance in the measured probabilities.

To calculate the expected runtime scaling, we first calculate the expected number of shots required to have a 99\% probability of obtaining a state with at least 0.8 approximation measure, then multiply that by the runtime of each shot. For $K$ independent shots of a quantum circuit, the total probability of measuring a state at least once is 
\begin{equation}
    p_{\rm total} = 1 - (1 - p_{\rm single})^K,
\end{equation}
where \(p_{\rm single}\) is the probability of measuring that state in a single shot. To include all states with approximation measure greater than 0.8 we take the sum total of all state probabilities. The number of shots required to have a 99\% probability of outputting such a state can be obtained by solving the inequality \(p_{\rm total} \ge 0.99\) for \(K\), 
\begin{equation}
    K_{99} \ge \frac{\log( 1 - 0.99 )}{\log(1-p_{\rm single})}.
\end{equation}
The estimated circuit execution runtime is then 
\begin{equation}
    T = t_{\rm single} \times K_{\rm min},
\end{equation}
where \(t_{\rm single}\) is the time required for a single shot. We compared these estimated runtimes with the recorded runtimes of the Gurobi algorithm which were executed on a single CPU core with a clock speed of 3.10GHz. We normalised each metric with respect to the minimal observed value which always occurred for the smallest trivial instance in order to obtain a metric that is agnostic of hardware speeds. In other words, we took the ratio \(T/T_0\) of the estimated runtime \(T\) against the runtime \(T_0\) of the easiest, smallest instance. For the quantum heuristics, the ratio clearly becomes independent of \(t_{\rm single}\) due to the proportionality assumption. However, note that using this metric is still not a fair comparison between classical and quantum methods, in particular since the Gurobi solution was always found to be optimal with an exact approximation measure of 1 for such small instances. In contrast, the quantum heuristics only had meaningful values when we accepted approximate solutions with more than 0.8 approximation measure. 
Moreover, this is not a direct comparison as the quantum circuits provide only an \emph{estimated} runtime from the observed probabilities but in reality the runtime would depend on a multitude of factors which were not considered in the experiments. In particular, the unpredictable nature of the classical optimisation loop meant that it is difficult to estimate a runtime for general instances. On the other hand, the \emph{exact} runtime of the classical Gurobi solver can be directly measured from input to output and is a good indication for other like instances. Nevertheless, we use the exact runtime scaling of Gurobi to see how the quantum heuristics compared. The runtimes are plotted in Fig \ref{fig:combined_2}(b). We note that each datapoint for CF-maQAOA has a large variance in the runtime ratio. These are likely due to a variety of factors, most likely attributed to the removal of RZZ gates, the optimisation of additional variational parameters, and the reduced noise in the circuits. All of these affect the final results in highly non-trivial ways. For further analysis, one could reduce the impact of these experimental errors by either increasing the number of shots in each quantum circuit execution, adding to the number of independent experiments by generating more problem instances, or repeating the same experiment over multiple time periods. Gurobi had the most consistent trend in runtime ratio with unsurprisingly an exponential growth for increasing numbers of cars. QAOA appeared to scale similarly to Gurobi although often an order of magnitude higher. The runtime ratio for CF-QAOA performed consistently lower than for QAOA. Moreover, the scaling seemed to compete closely with that of Gurobi. The runtime ratio for CF-maQAOA performed better than other methods although is not consistent.

It is still too early to conclude whether the connectivity-forced heuristic can compete with classical algorithms like Gurobi in the long run. However, we observed that it is possible to make changes to quantum circuits in non-intuitive ways that can bring about better performance closer to that of Gurobi. It is still unclear how much entanglement is lost when removing two-qubit gates in the connectivity-forced QAOA variants. Further work could analyze the entanglement properties of such circuits, as well as its classical simulability. In another direction, it is possible to devise a better compression strategy for the ansatz. For instance, incorporating the problem constraints directly into the mixer and initial states could drastically reduce the number of terms in the Hamiltonian, reducing the number of two-qubit gates

\section{Conclusion}

In our numerical and device experiments, we investigated the performance of the Quantum Approximate Optimization Algorithm (QAOA) on traffic problems constructed as Quadratic Unconstrained Binary Optimisation (QUBO) problems. Classical simulations were used to investigate QAOA performance on traffic problems under different parameter initializations. In our numerical experiment, the Trotterized Quantum Annealing (TQA) initialization outperformed random initialization (RI). We observed that QAOA with precomputed parameters gives high-quality solutions to traffic instances, a result known to hold for MaxCut instances~\cite{Brando2018ForFC, fixed-angle-QAOA, transfer_median, Galda2021TransferabilityOO, QAOA-ML}.

On physical hardware, circuit noise destructively impacted the performance due to deep circuits after compilation. We introduced a connectivity-forced circuit compression heuristic (CF-QAOA) whereby long-range RZZ gates over non-connected qubits on the hardware are removed. Compared to QAOA, this heuristic improved approximation measures of the final solution state when run on physical quantum devices. Introducing extra variational parameters into the subsequent connectivity-forced QAOA circuits (CF-maQAOA) provided a trade-off between increased performance and additional classical optimization. We observed that the expected runtime scaling of QAOA and CF-QAOA scaled similarly to Gurobi. Compared to QAOA, the expected runtime scaling on physical quantum hardware was lower for CF-QAOA and even lower for CF-maQAOA, although with a higher variance. Compared to baseline QAOA, while CF-QAOA does worse in noise-free simulations, it has higher performance on physical hardware. With additional classical finetuning, the CF-maQAOA variant provides a potential improvement to CF-QAOA in both performance and runtime scaling.

\begin{acknowledgments}
This work was supported by the University of Melbourne through the establishment of an IBM Quantum Network Hub at the University. The authors also acknowledge support from the Ford Alliance Program and the University of Melbourne’s Zero Emission Energy Laboratory (ZEE Lab) funded under the Victorian Higher Education State Investment Fund (VHESIF).
\end{acknowledgments}

\bibliography{main}

\end{document}